\documentclass[pre,preprint]{revtex4}
\usepackage{epsfig,epsf}
\draft
\begin{document}
\title{Information Entropy and Correlations in Prime Numbers}

\author{Pradeep~Kumar, Plamen~Ch.~Ivanov, H.~Eugene Stanley} 

\address{
Center for Polymer Studies and Department of Physics,
  Boston University, Boston, MA 02215\\
}

\date{\today}

\begin{abstract}
 
The difference between two consecutive prime numbers is called the
distance between the primes. We study the statistical properties of the
distances and their increments (the difference between two consecutive
distances) for a sequence comprising the first $5\times 10^7$ prime
numbers. We find that the histogram of the increments follows an
exponential distribution with superposed periodic behavior of
period three, similar to previously-reported period six
oscillations for the distances.

\end{abstract}

\maketitle


Recent reports indicate that many physical and biological systems
exhibit patterns where prime numbers play an important role. Examples
range from the periodic orbits of a system in quantum chaos to the life
cycles of species \cite{BJ2,BJ5,MP1,CMK1,SRD1,BJ3,BJ4}. Recent work
reports on a potential for which the quantum energy levels of a particle
can be mapped onto the sequence of primes \cite{GM1}. Furthermore, it
has been shown that a gas of independent bosons with energies equal to the
logarithm of consecutive primes possesses a canonical partition function
coinciding with the Riemann function \cite{BJ1}. The partition function
of a system with energies equal to the distances between two consecutive
prime numbers behaves like a set of non-interacting harmonic oscillators
\cite{M1}. Most recently, power-law behavior in the distribution of
primes and correlations in prime numbers have been found \cite{M2},
along with multifractal features in the distances between consecutive primes
\cite{M3}. Previous work thus further motivates studies
of prime numbers using methods of statistical physics. Here, we focus on
the statistical properties of the distances between consecutive prime
numbers and the increments in these distances [Fig.~1].



Since the distribution of distances is well-studied, we discuss the
occurrence frequency of increments between consecutive distances. We
find that the distribution of increments [Fig.~2(a)] exhibits large
peaks for given values of the increments and medium and small peaks for
other values, and that these peaks follow period-three
oscillation. Specifically, we find that the increments with values of
$6k+2$ ($k=0,1,2,3...$) have the highest occurrence frequency, followed
by increments with values of $6k+4$. Values of $6k$ are relatively rare
and correspond to the small peaks in the distribution. This regularity
is present for both positive and negative increments and does not depend
on the length $N_p$ of the sequence. We also find that the occurrence
frequency of increments decreases exponentially and that this
exponential behavior is well pronounced for both large and small peaks,
forming a ``double-tent'' shape [Fig.~2(b)].

We find exponential behavior with superposed periodic behavior with
period-three oscillation for the distribution of increments similar to
the period-six oscillation for the distribution of distances
\cite{M1}. Further, we find that the occurrence frequency of a positive
increment is almost the same as the occurrence frequency of its negative
counterpart for a given sequence length $N_p$ [Fig.~2(c)].


In summary, we find a new statistical feature in the sequence of
increments between consecutive prime distances. We find a period-three
oscillation in the distribution of increments and this distribution
follows an exponential form.

This empirical observation may be of importance in further
understanding the nature of prime numbers as well as those physical and
biological processes where prime numbers play a role.

\acknowledgments We thank M.~Wolf,  S.~Havlin and M.~Taqqu for helpful discussions.

\pagebreak


\begin{figure}[h]
\begin{center}
\epsfxsize=3in \epsffile{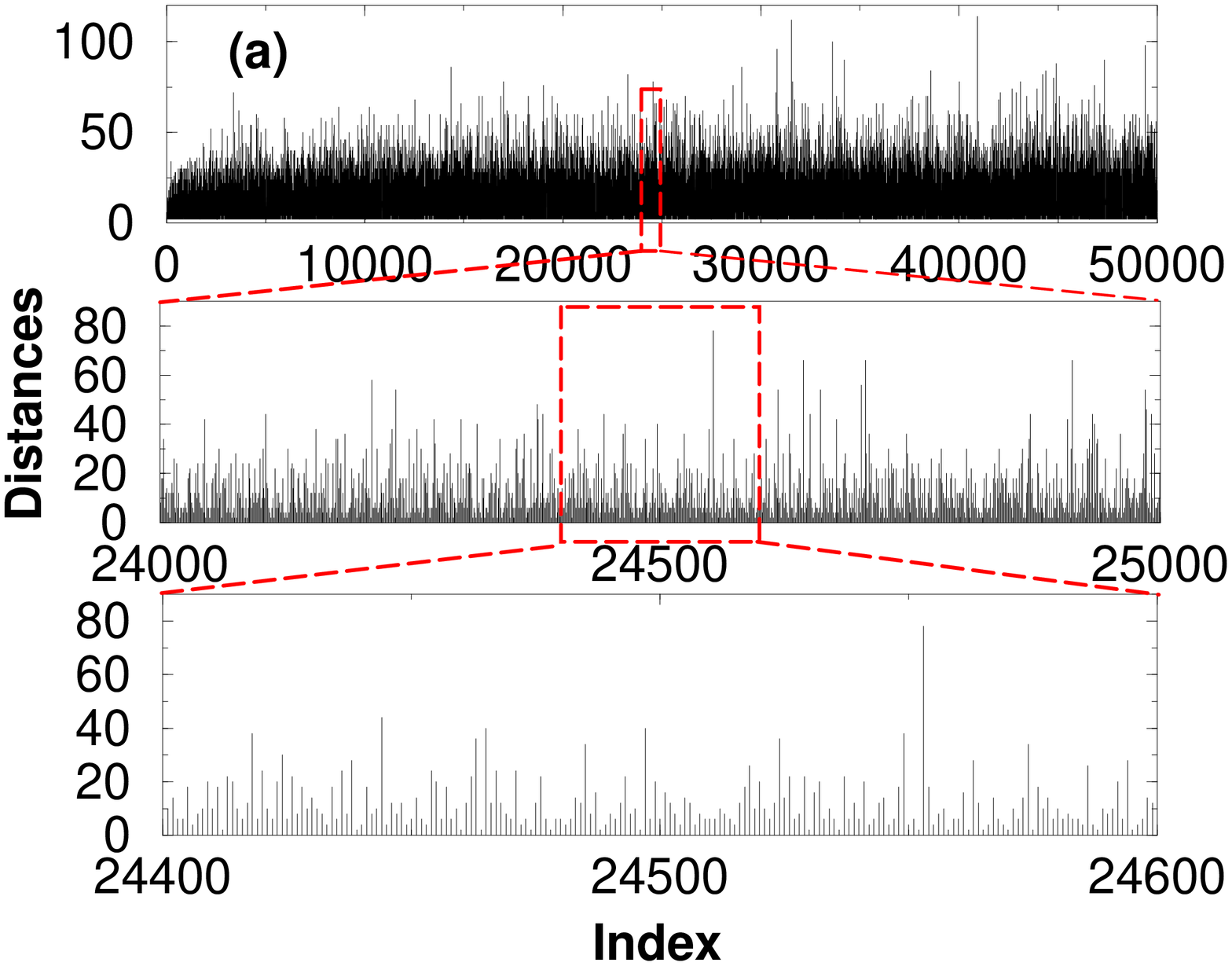} \hspace*{0.5cm} \epsfxsize=3in
\epsffile{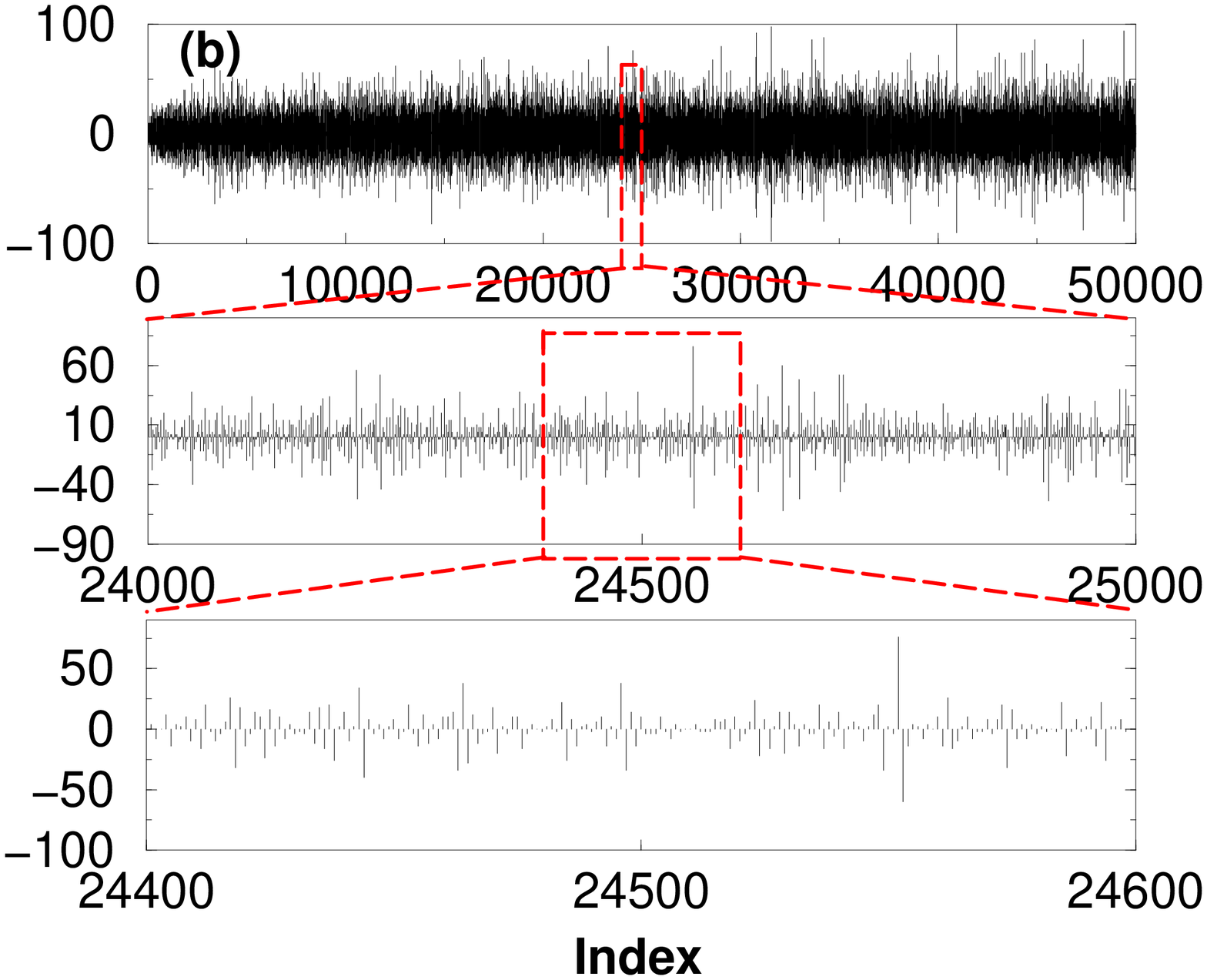}
\end{center}
\caption{ Distances between consecutive prime numbers (indexed
  sequentially) and their increments.  (a) The first $5\times 10^4$
  distances between consecutive prime numbers. (b) The first $5\times
  10^4$ increments.}
\end{figure}

\begin{figure}
\begin{center}
\epsfxsize=3in \epsffile{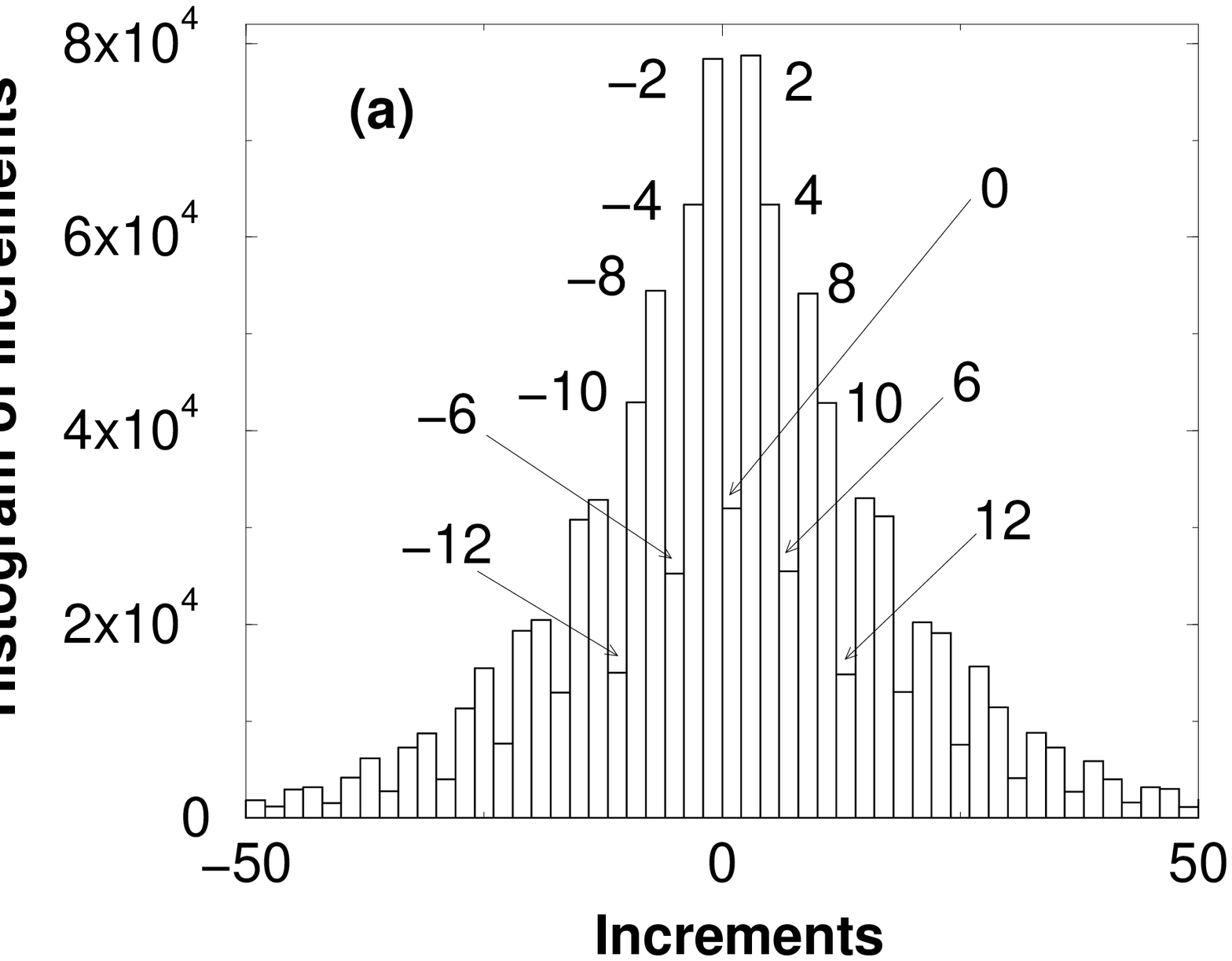}  \epsfxsize=3in \epsffile{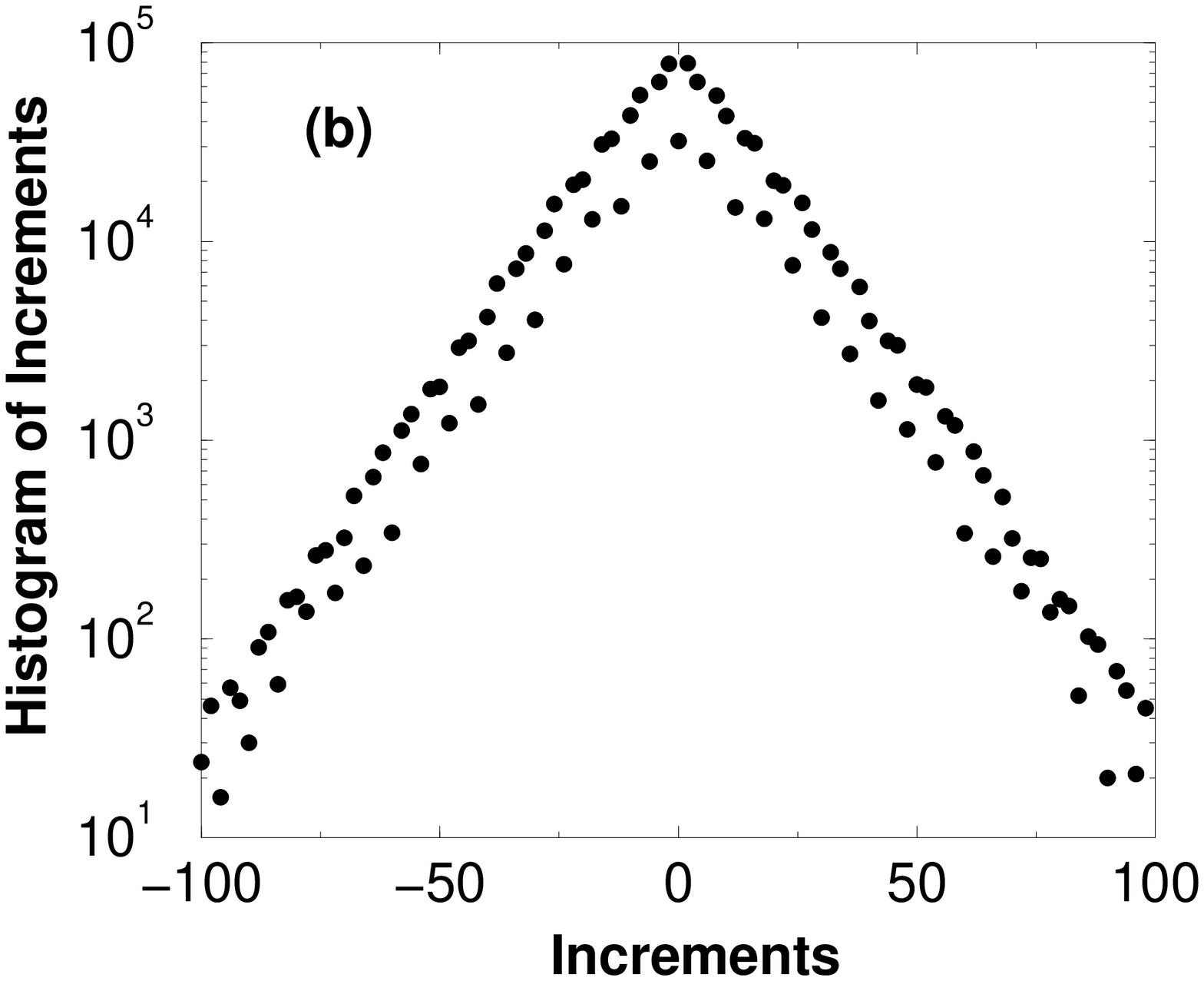}
\epsfxsize=3in \epsffile{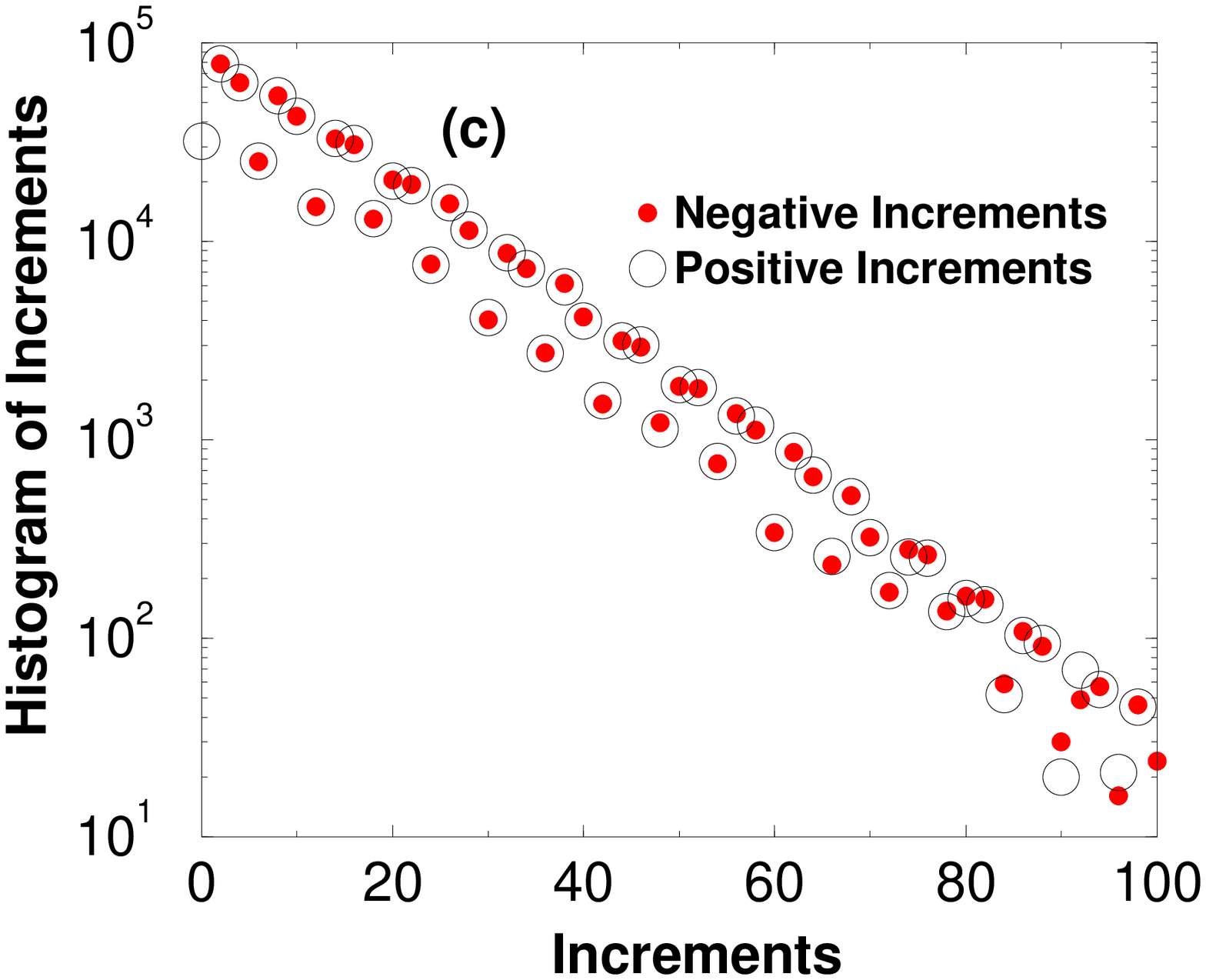}
\end{center}
\vspace*{-0.5cm}
\caption{(a) Histogram of increments in the distances between
consecutive prime numbers for the sequence of the first $N_p = 10^6$
primes. The bin width is 1. The occurrence frequency of increments with
given values exhibits a robust period-three oscillation. Increments with
values $\pm(6k+2)$ ($k = 0,1,2,3,....$) occur most often, increments
with values $\pm(6k+4)$ occur less often, and increments with values
$\pm6k$ are rare.  This regularity is always present regardless of the
sequence length $N_p$.  (b) Tent-shape of the histogram of increments on
a linear-log plot suggests an exponential form for large, medium, and
small peaks. The top curve (corresponding to large and medium peaks) is
thus $\pm2,\pm4,\pm8,\pm10,\pm14,\pm16....$, while the bottom curve
(corresponding to small peaks) is $0,\pm6,\pm12,\pm18,....$.(c) Symmetry
is observed in the frequency of occurrence of positive increments and the
corresponding negative increments (see also (a)).} 
\end{figure}








\end{document}